\documentclass[preprint,authoryear,12pt]{elsarticle}
\usepackage{amsmath,amssymb,epsfig,amsfonts,epsfig,graphicx}

\usepackage{epsfig}
\usepackage{epstopdf}
\usepackage{amssymb}
\usepackage{amsmath} \usepackage{stmaryrd}
\usepackage{epic} \usepackage{eepic}
\usepackage{latexsym}
\usepackage{fancyhdr}
\usepackage[bookmarks=false]{hyperref}
\renewcommand{\phi}{\varphi}

\newcommand{\beq}{\begin{equation}}
\newcommand{\eeq}{\end{equation}}
\newcommand{\bea}{\begin{eqnarray}}
\newcommand{\eea}{\end{eqnarray}}
\newcommand{\beas}{\begin{eqnarray*}}
\newcommand{\eeas}{\end{eqnarray*}}

\usepackage{color}

\newcommand{\cred}{}

\newcommand{\crev}{}

\makeatletter
\def\ps@pprintTitle{%
\let\@oddhead\@empty
\let\@evenhead\@empty
\let\@oddfoot\@empty
\let\@evenfoot\@oddfoot
}
\makeatother

\graphicspath{ {figures/} }

\begin{document}


\begin{frontmatter}

\title{On the additive manufacturing, post-tensioning and testing of bi-material tensegrity structures}

\author{A.~Amendola}
\ead{adaamendola1@unisa.it}
\address{Department of Civil Engineering, University of Salerno,84084 Fisciano(SA), Italy}

\author{E. Hern\'andez-Nava, R.~Goodall, I.~Todd}
\ead{mtq10eh@sheffield.ac.uk, r.goodall@sheffield.ac.uk, i.todd@sheffield.ac.uk}
\address{Department of Materials Science and Engineering, University of Sheffield, Mappin Street, Sheffield, S1 3JD, UK}

\author{R.E.~Skelton}
\ead{bobskelton@ucsd.edu}
\address{University of California San Diego, MAE/Aero, 9500 Gilman Dr., La Jolla, CA 92093, USA}

\author{F.~Fraternali}
\ead{f.fraternali@unisa.it}
\address{Department of Civil Engineering, University of Salerno,84084 Fisciano(SA), Italy}

\begin{abstract}

{{An investigation on the additive manufacturing and the experimental testing of 3D models of tensegrity prisms and columns is presented. An electron beam melting facility (Arcam EBM S12) is employed to 3D print structures composed of tensegrity prisms endowed with rigid bases and temporary supports, which are made out of the titanium alloy Ti6Al4V. The temporary supports are removed after the additive manufacturing phase, when Spectra cross-strings are added to the 3D printed models, and a suitable state of internal prestress is applied to the structure.
The experimental part of the study shows that the examined structures feature sitffening-type elastic response under large or moderately large axial strains induced by compressive loading. Such a geometrically nonlinear behavior confirms previous theoretical results available in the literature, and paves the way to the use of tensegrity prisms and columns as innovative mechanical metamaterials and smart devices.
 }}

\end{abstract}


\end{frontmatter} 

\medskip

\section{Introduction}\label{intro}

One of the most common  techniques nowadays employed for the fabrication and the rapid prototyping of innovative periodic structures and lattice materials is additive manufacturing (AM). Several fabrication methods have been proposed in this field, with resolution ranging from the centimetre- to the nanometre-scale. Notable methods include polyjet 3d printing technologies; Electron Beam Melting (EBM);  x-ray lithography; deep UltraViolet lithography; soft lithography; two-photon polymerization; atomic layer deposition; and projection micro-stereolitography, among other available methods (refer to \cite{Buckmann:2014}, \cite{Maldovan:2013}, \cite{Meza:2014}, \cite{Sheffield}, \cite{Zheng12}, and references therein). 
The processing of detailed components through the consecutive addition of small quantities of material, usually in layers thinner than 100 $\mu$m, makes AM capable of producing highly complex structures at different scales. 
EBM is a popular AM technique used to fabricate metallic structures. Its name comes from how materials are deposited in the technique, in which, it makes use of a beam of electrons to fully melt powder particles of a conductive material. This melting operation is performed in a selective way through a surface file, commonly STL. The use of EBM has been successfully demonstrated in  several studies of structured materials as; lattices with graded porosity \citep{Sheffield}, metallic foams \citep{Hernandez}, optimized topologies \cite{Smith} and structures with negative Poisson's ratio \citep{Cormier} to name a few.

Tensegrity structures are axially loaded prestressable structures. Motivated by nature, where tensegrity concepts recurrently appear \citep{Skelton2010c}, 
engineers have only recently developed efficient analytical methods to exploit tensegrity concepts in engineering design. 
Traditional construction of tensegrity structures through manually assembling of different components can represent a challenging task. Potentially resulting in a series of mitigated results because of structural and connecting issues; eccentricities at the joints, unwanted movements and friction (refer, e.g., to \cite{prot} and references therein). Hence in order to systematically study complex tensegrity systems, the development of fabrication techniques based on AM would be of great help. 
This paper proposes for the first time a method to construct tensegrity systems in a two-step process, using AM to generate a tensegrity structure without its tension members (instead an artificial set of members to stabilize the structure, prior to adding the prestress, and secondly to remove the artificial members and add tensile members after the AM step is finished.
 The present work investigates the quasi-static response under axial compressive loading of bi-material models of tensegrity structure partially produced through EBM. 

The paper begins by describing the Computer Aided Design (CAD) of the structures to be manufactured,
which include one or more building blocks (tensegrity prisms) formed by three titanium alloy (Ti6Al4V) bars; two Ti6Al4V plates (bases), and three Spectra fibers connecting the base plates (Sect. \ref{Design}).
Sect. \ref{materials} describes the EBM process used for the AM manufacturing of intermediate models without the Spectra fibers, 
while Sect. \ref{post-tension} illustrates the placement and the post-tensioning of such elements within the final structures.
The experimental part of the study presents quasi-static compression tests that investigate on the response of the examined  structures under compressive axial loading.
The given results validate the theoretical predictions presented in \citep{prot, JMPS14}, highlighting the geometrically non-linear nature of the axial force vs. axial displacement response of such structures (Sect. \ref{comptest}). We end in Sect. \ref{conclusion}, by drawing the main conclusions of the present study and future research lines.

\section{Geometric models of tensegrity prisms and columns}\label{Design}

We focus in the present Section on the computer aided design of provisional models of structures showing tensegrity prisms endowed with rigid bases as building blocks \citep{prot}. Each prism is composed of three struts, two rigid bases (a top-base, and a bottom-base), and three cross-cables. 

Figs. \ref{Ti_model}, \ref{Ti_model_chain}  show a computer models of {metallic structures} formed by the struts and the terminal bases of the building blocks, plus some sacrificial supports  connecting the bases of the building blocks (vertical elements in Figs. \ref{Ti_model}, \ref{Ti_model_chain}). The connections between the bars and the bases are realized through hemispherical (terminal blocks) or spherical (internal blocks) joints (cf. Figs. \ref{Ti_model}, \ref{Ti_model_chain}). Such provisional structures will be additively manufactured in a titanium alloy (Ti6Al4V) through the EBM facility Arcam S12 (cf. Sect. \ref{materials}). Next, the 3D printed structures will be finished with the addition of tensioned Spectra strings connecting the struts and the terminal bases of the building blocks. Once the insertion of the Spectra strings will be completed, and a suitable prestress will be applied to the structure, the sacrificial supports will be removed (\textit{post-tensioning approach}, cf. Sect.\ref{post-tension}). 

We now show in detail how we design the building blocks of the structures under investigation.
To ensure zero bending moments, the extremities of the bars composing such units should have almost zero diameter at the nodes.
Unfortunately, it is impossible to additively manufacture elements with nearly zero diameter, due to the limited precision tolerance of the available printing machine.
Indeed, the  ARCAM S12 EBM facility allows to smoothly manufacture features with size down to 0.4 mm \citep{Sheffield}.
Such a limitation also affects the manufacturing of the tensile elements (cables or strings) of the building blocks, due to the small diameter that is needed for such elements.
As explained previously, our fabrication strategy makes use of 3D printed temporary models in which the strings are replaced by sacrificial struts with diameter $d^{\ast}=0.5$ mm. The latter are removed in a second phase, when Spectra strings are manually inserted into the building blocks, with the aim of `sewing' the 3D printed models (Sect.\ref{post-tension}).
To this end, the models to be 3D printed show holes passing through the joints, which will host the Spectra strings in the post-tensioning phase. 
We design the compressive elements (bars or struts) of the temporary models with a bi-conic shape obtained by joining two truncated cones. The latter feature decreasing radii towards the extremities (Fig. \ref{Ti_model}), so as to minimize the bending rigidity of the nodes. Let $d$ denote the minimum diameter of the struts (at the extremities), and let $D$ denote the maximum diameter of such elements (at the mid-span).
Following {\cite{AIP12} we make use of the following tapering ratio: $\beta = D/d = 6$.

Overall, each {building block} to be 3D printed is composed of three truncated bi-cones (bars), three tacrificial supports, two triangular plates of thickness $t$ (bases), and six hemispherical or spherical joints with radius $r$ (Fig. \ref{Ti_model}).
Table \ref{CAD} shows the geometric properties of such elements, making use of the symbols $s_N$, $b_N$ and $h_N$ to respectively denote the length of the edges of the base plates; the length of the struts; and the height of the prism (measured from the centre to centre of the bases), in correspondence with the unstressed (or natural) configuration of the structure.
Fig. \ref{Ti_model_chain} shows temporary models of tensegrity columns obtained by superimposing different numbers of building blocks. Each block can be either `right-handed' or `left-handed', depending whether the upper base is clockwise or counter-clockwise twisted with respect to the lower base, respectively. The examined columns can stack blocks with the same orientation (Fig. \ref{Ti_model_chain}x-y), or blocks with different orientations (Fig. \ref{Ti_model_chain}z-w).

\begin{figure}[hbt] \begin{center}
\includegraphics[width=6cm]{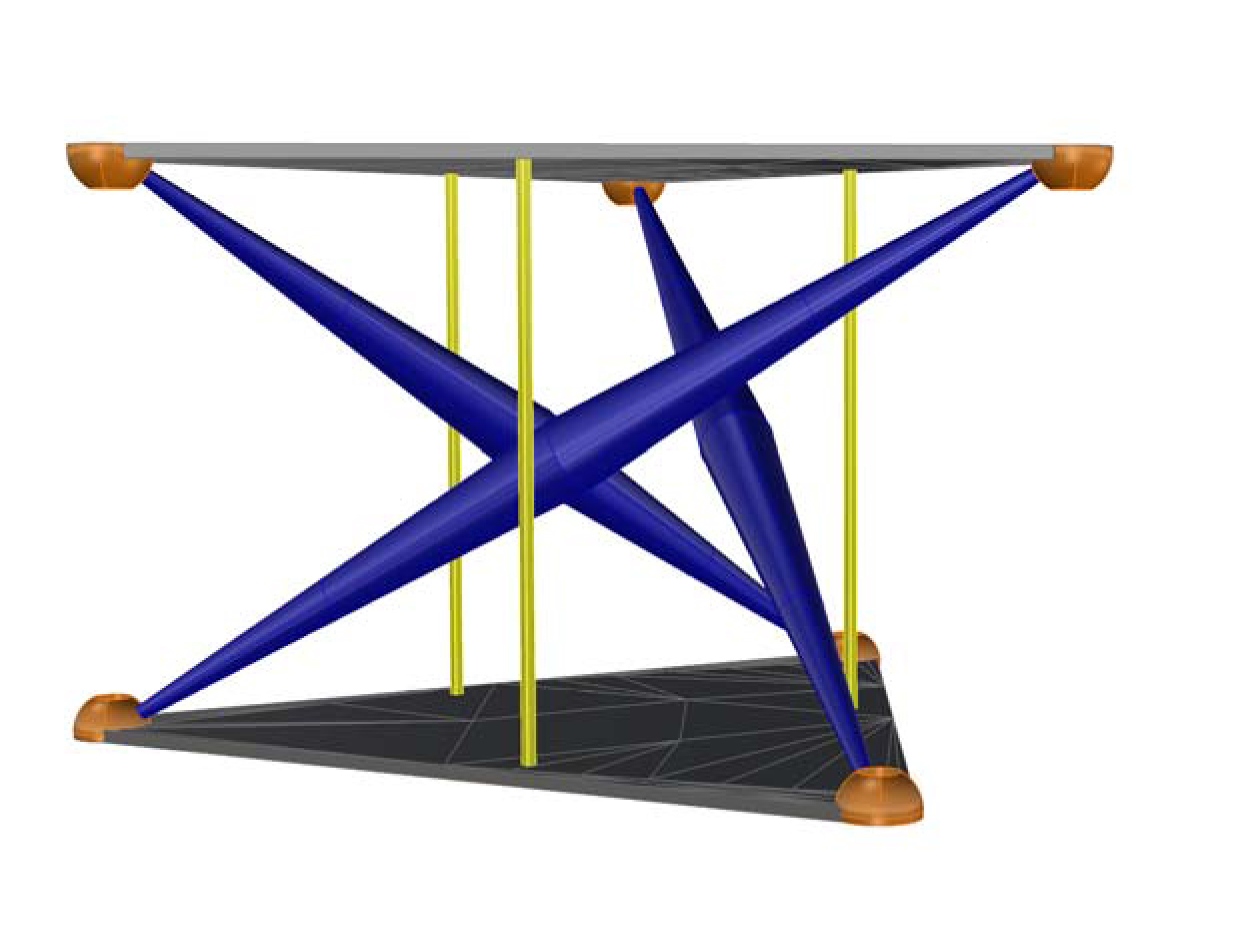}
\caption{Graphical model of a {building block} composed of three sacrificial supports (yellow), six semi-spherical joints (orange), three struts (blue) and two terminal plates (gray).}
\label{Ti_model}
\end{center}
\end{figure}

\begin{figure}[hbt] \begin{center}
\includegraphics[width=8cm]{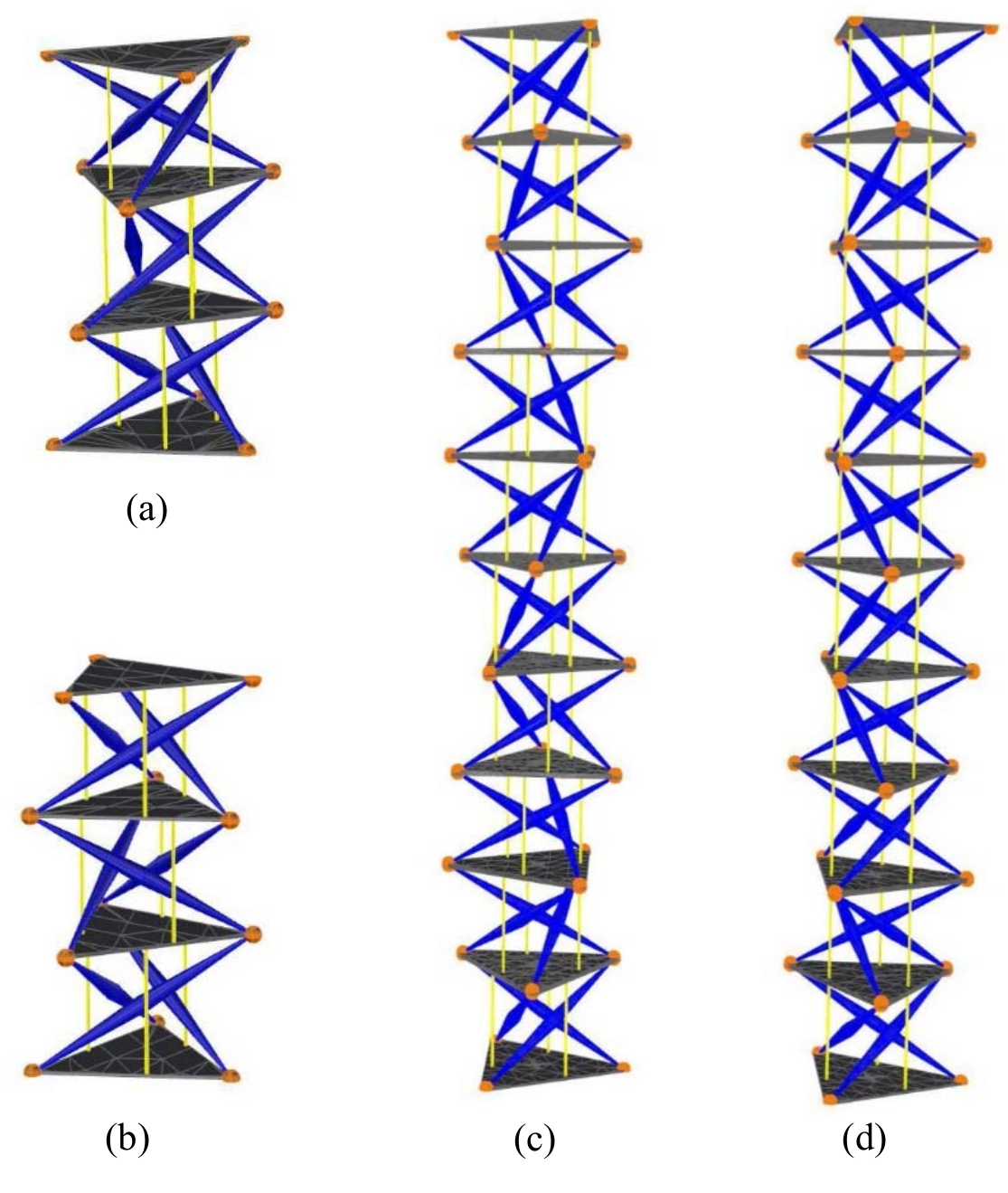}
\caption{Collection of tensegrity columns to be 3D printed using different numbers of prisms and different orientations: (a) column with three left-handed prisms, (b) column with three left-right-left-handed prisms, (c) column with ten left-handed prisms, (d) column stacking ten prisms with different orientations (alternating left- and right-handed prisms).}
\label{Ti_model_chain}
\end{center}
\end{figure}

\begin{table}[htbp]
	\centering
				\begin{tabular}{| c | c | c | c |c | c | c | c | c |}
    \hline
 $t$ (mm) &$r$  (mm)  & $s_N$ (mm) & $b_N$ (mm) &$D$(mm)&$d$ (mm) & $d^{\ast}$ (mm) & $h_N$ (mm) \\ \hline
 $1.0$& 1.5 & 34.08 &44.06 & 3.0 & 0.5 & 0.5 &22.29 \\ \hline
			\end{tabular}
	\caption{Geometrical properties of the building blocks of the examined models}
		\label{CAD}
\end{table}

\section{Additive manufacturing of physical models} \label{materials}
The present section focuses on the EBM technique employed to process the CAD models illustrated in the previous section.
We begin by describing the geometrical and chemical characteristics of the employed Ti6Al4V alloy Next, we briefly describe the employed manuacturing process.
We end by presenting an experimental study on the influence of the direction in which the layers are built up in the EBM (the build direction) on the mechanical properties of the melted Ti-6Al-4V alloy.

\subsection{Raw materials}\label{rawmaterials}
The  manufacturing process examined in the present work starts with the spreading of a thin layer of Ti4Al6V prealloyed powder into the build platform of the printing machine (Arcam EBM S12). Such a powder is provided by the manufacturer and is made of spherical particles with 45 $-$ 100 $\mu$m diameter.
The raw material Grade 5 is under the standards of the American Society of Testing and Materials (ASTM Grade 5, 6al-4V) \citep{ASTM14}.
It must be considered, however, that the continuous material reuse, which is a common practice in the EBM manufacturing, might slightly change the chemical composition of the Ti6Al4V powder (for example by gradual pick up of oxyge). Tab. \ref{chemistry} provides the chemical composition experimentally determined for the powder employed in the previous study.

\begin{table}[ht]\label{chemistry}
\caption{Chemical composition of the Ti-6Al-4V prealloyed powder in wt\%.}
\centering
\begin{tabular}{ccccccc}
\hline
Ti & Al & V & Fe & C & N & O \\ [0.5ex]
\hline
88.28 & 6.88 & 4.27 & 0.18 & 0.007 & 0.026 & 0.33 \\
\end{tabular}
\label{table:themes}
\end{table}

\subsection{Manufacturing process}\label{AM}
EBM works in vacuum conditions at $1.9\times 10^{-3}$ mbar, and heats the base of the build area up to the temperature of 720{$\circ$}C. 
Layers of Ti-6Al-4V powder are progressively deposited, heated and melted, according to a sliced version of the CAD model to be manufactured, which is prepared by an internal preprocessor of the Arcam EBM S12. Layer heating (using a defocused beam) is employed to reduce the energy needed for the melting phase. 
The EBM working parameters are varied for the bars, joints and plates, with the of ensuring uniform material density throughout the model, as much as possible (cf. Tab, \ref{parameters}). The voltage is kept constant at 60kV. Once manufactured, all remnants of unmelted powder are removed using compressed air.

\begin{table}[ht]
\caption{Beam parameters employed for the manufacturing of the model parts.}
\centering
\begin{tabular}{cccc}
\hline
& Beam current & Beam speed & Focus Offset \\ [0.5ex]
&(mA) & (mm/s) & \\
\hline
\quad Preheat & 30 & 14600 & 50 \\
\quad Plates & 17 & 500 & 19 \\
\quad Spheres &1.7 & 200 &0  \\
\quad Bars & 1.7 & 200 & 0  \\
\end{tabular}
\label{parameters}
\end{table}

\subsection{Parent material properties}

We printed twelve tensegrity prisms with two different building orientations for material characterization purposes.
The min goal of such a study was to detect the influence of the EBM angle deposition on the mechanical properties of the melted Ti-6Al-4V alloy.
We built six samples along the vertical direction (V specimens: prism bases parallel to the building plate), and  six other samples along the horizontal direction (H specimens: prism bases perpendicular to the building plate).
Parts of such specimens, manually cut using shears, were subjected to
tensile tests according to the ASTM standard E8-13a \citep{ASTM13a}.
The data obtained from tensile tests include the 0.2 \% Yield strength (YS), Ultimate Tensile Strength (UTS), and Young's modulus (YM), see Table \ref{table:properties}.
The yield strength of V- and H-specimens was found slightly different (V-specimens more resistant than H-specimens, cf. Table \ref{table:properties}), mainly due to changes in material porosity build direction \citep{Murr2010a}. It is worth mentioning that the replicability of material properties in the AM of  lattice materials in general is subject of debate. This is as the fraction of defects, such as internal porosity and surface roughness, dramatically affect the material strength, especially in the case of small features, such as, e.g.,  the nodal junctions \citep{Hernandez}.
The results presented in Table \ref{table:properties} led us to prefer the vertical deposition technique for the manufacturing of the final specimens (Figs. \ref{EBM_prism}, \ref{EBM_3LR_chain} and \ref{EBM_10LR_chain}).

\begin{table}[ht]
 \caption{Bulk material properties based on tensile tests for two build orientations.}
 \centering
 \begin{tabular}{cccc}
 \hline
Build orientation & 0.2 \% Y.S. & UTS & Young's modulus \\ [0.5ex]
 \hline
  & (MPa) & (MPa) & (GPa) \\
Horizontal & 881$\pm$ 8.4 & 1040$\pm$6.7 & 110$\pm$2.1\\
Vertical & 923$\pm$10.7 & 1042$\pm$9.2 & 116$\pm$2.8 \\
 \end{tabular}
 \label{table:properties}
 \end{table}

\begin{figure}[hbt] \begin{center}
\includegraphics[width=8cm]{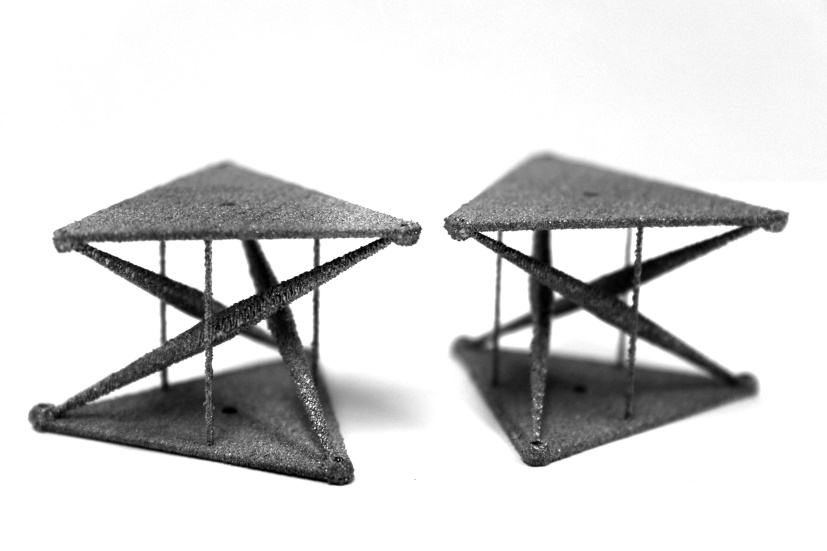}
\caption{EBM-printed left-handed (left) and right-handed (right) tensegrity prisms.}
\label{EBM_prism}
\end{center}
\end{figure}

\begin{figure}
\centering
\includegraphics[width=.5028\textwidth]{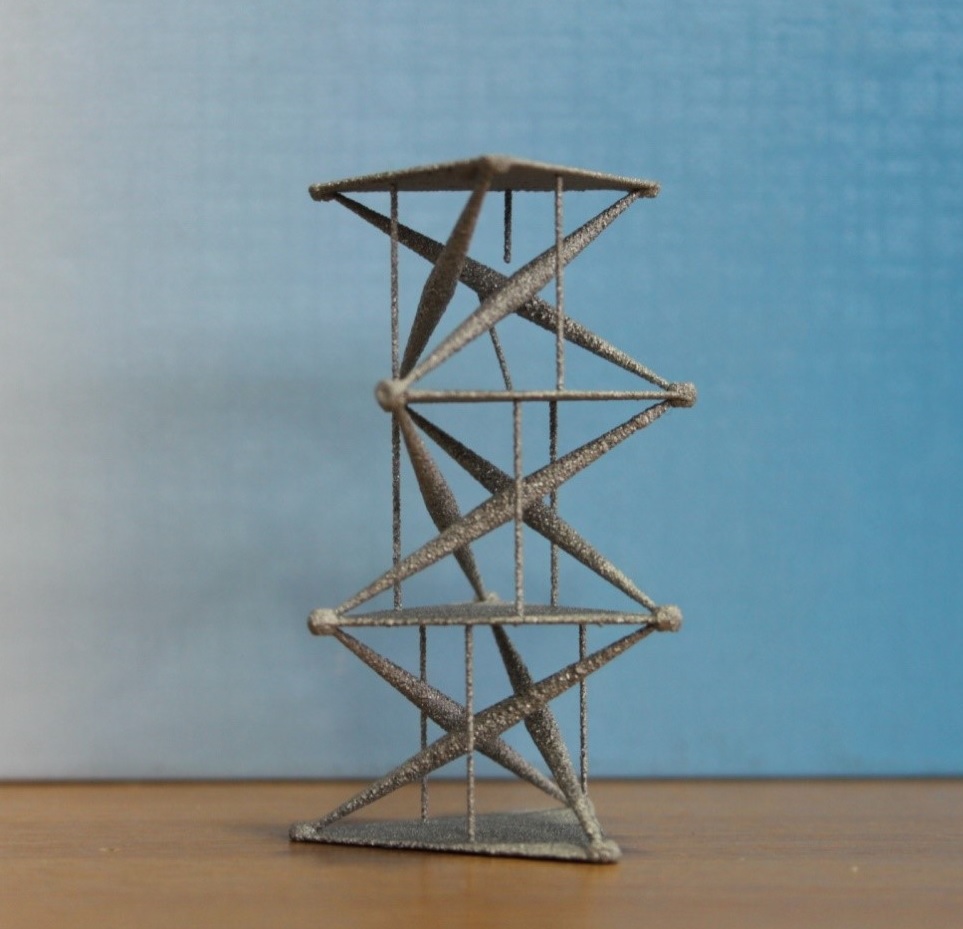}\hfil
\includegraphics[width=.49\textwidth]{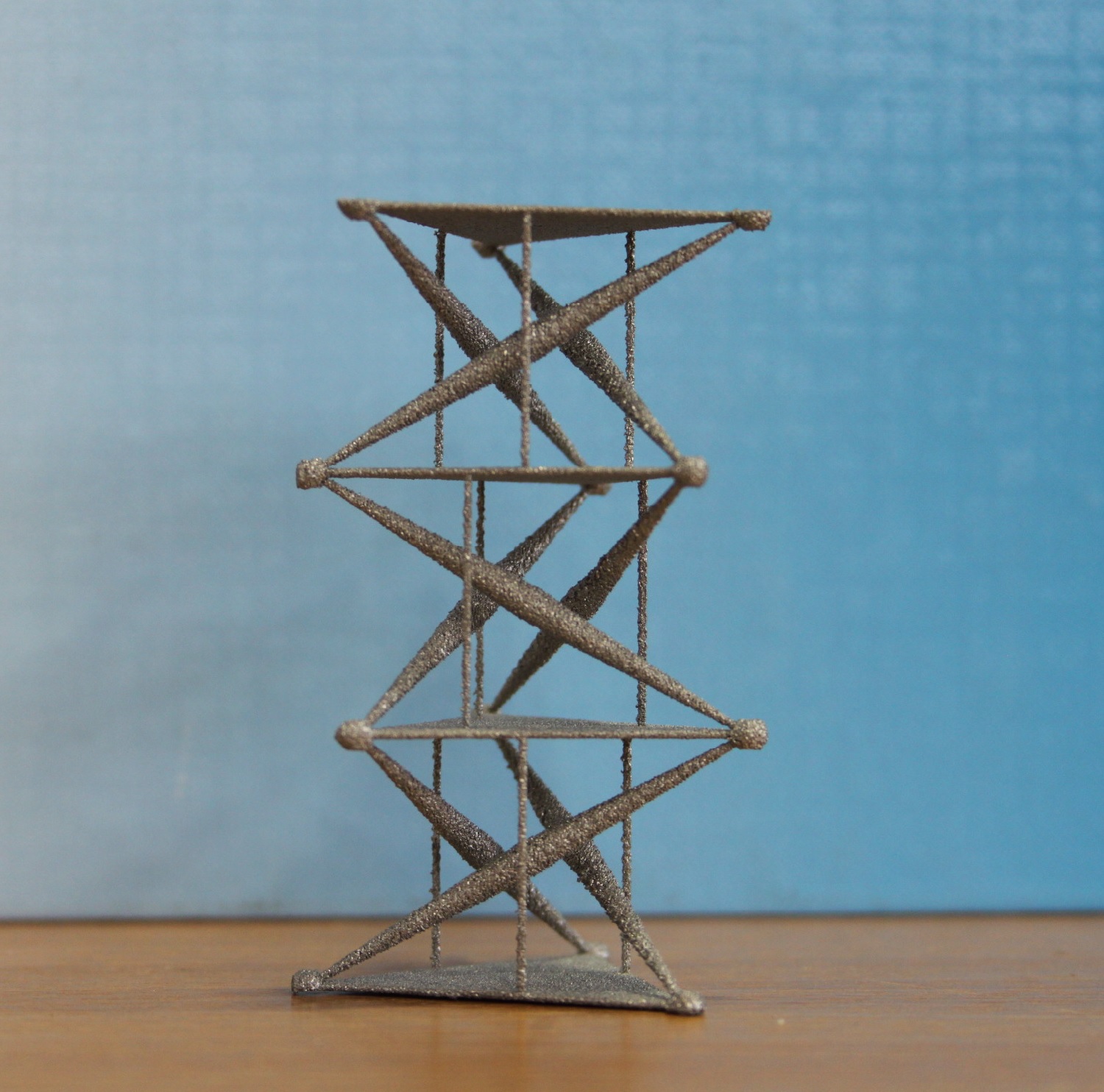}
\caption{EBM-printed columns with three left-handed prisms (left), and three left-right-left-handed prisms (right).}
\label{EBM_3LR_chain}
\end{figure}

\begin{figure}[hbt] \begin{center}
\includegraphics[width=8cm]{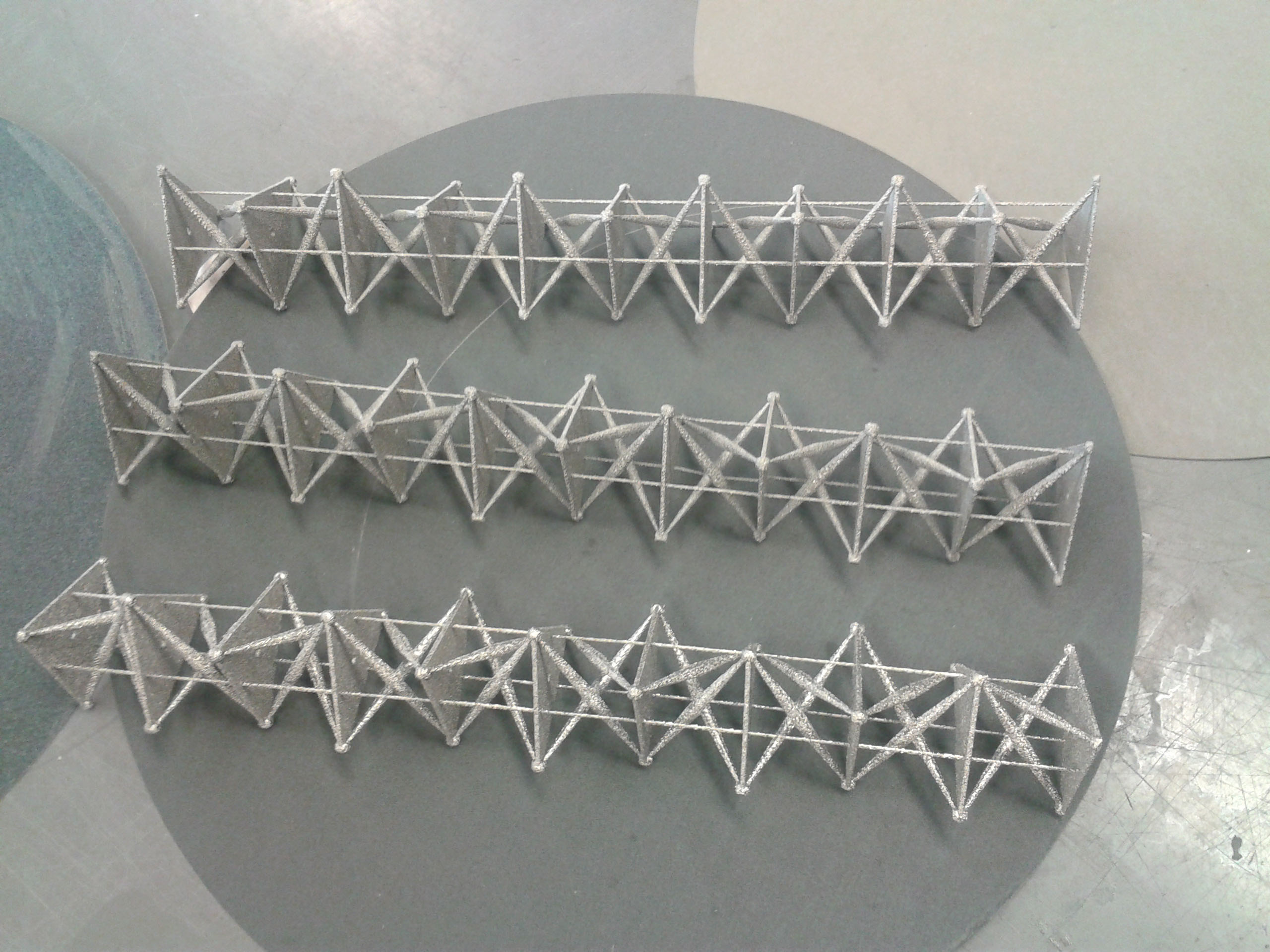}
\caption{EBM-printed columns with ten building blocks, obtained by alternating left-handed and right-handed prisms.}
\label{EBM_10LR_chain}
\end{center}
\end{figure}

\section{Post-tensioning of 3D printed specimens} \label{post-tension}
As anticipated, we `sewed' the 3D printed structures by slipping Spectra\textsuperscript{\textregistered} fibers with 0.1 mm diameter through the holes crossing the joints. We used Spectra fibers commercialized by Shimano American Corporation (Irvine, CA, USA), with declared maximum load supporting capacity of 5 kg. Hereafter, we assume that such string elements have Young modulus $E_s$=$5.48 GPa$, based on the results of the experimental study presented in \citep{prot}. 
Since the cross-section area of the strings is $A_s = \pi \ 0.1^2/ 4 = 0.0078 \ \mbox{mm}^2$, their axial stiffness is computed as: $E_s A_s = 0.043  \ \mbox{kN}$.
After slipping the Spectra strings through the holes crossing the joints, we fixed them in correspondence with the top-base of the structure through slipknots (Fig. \ref{Sewed_prisms}). 
Next, we removed the  sacrificial supports of the 3D printed structure (cf. Figs. \ref{EBM_prism}, \ref{EBM_3LR_chain} and \ref{EBM_10LR_chain}), and we tensioned the strings by fixing them to the bottom-base.
A level was placed on the top-base of the structure during the post-tensioning phase, in  order to ensure that such a base was horizontal, and the strings approximatively carried equal
tensile forces.
We applied the post-tension approach to manufacture single prisms, either left-handed (L-prisms, Fig.  \ref{Sewed_prisms}-left) or right-handed (R-prisms, Fig.  \ref{Sewed_prisms}-right); columns showing three left-handed prisms (3LLL-columns, Fig.  \ref{Sewed_columns}-left);
and columns showing two terminal left-handed prisms and one central right-handed prism (3LRL-columns, Fig.  \ref{Sewed_columns}-right). 
The application of the post-tensioning approach to larger structures (such as, e.g., the ten-prism columns shown in Fig. \ref{EBM_10LR_chain}) will be experimented in future work.

\begin{figure}[hbt] \begin{center}
\includegraphics[width=8cm]{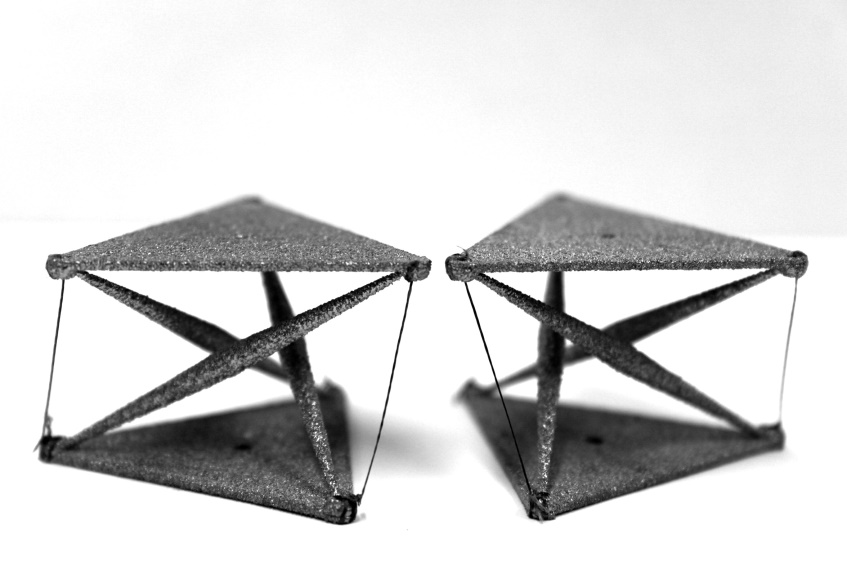}
\caption{L-prism (left) and R-prism (right)  samples obtained by `sewing' the EBM-printed models with Spectra strings.}
\label{Sewed_prisms}
\end{center}
\end{figure}

\begin{figure}[hbt] \begin{center}
\centering
\includegraphics[width=0.4\textwidth]{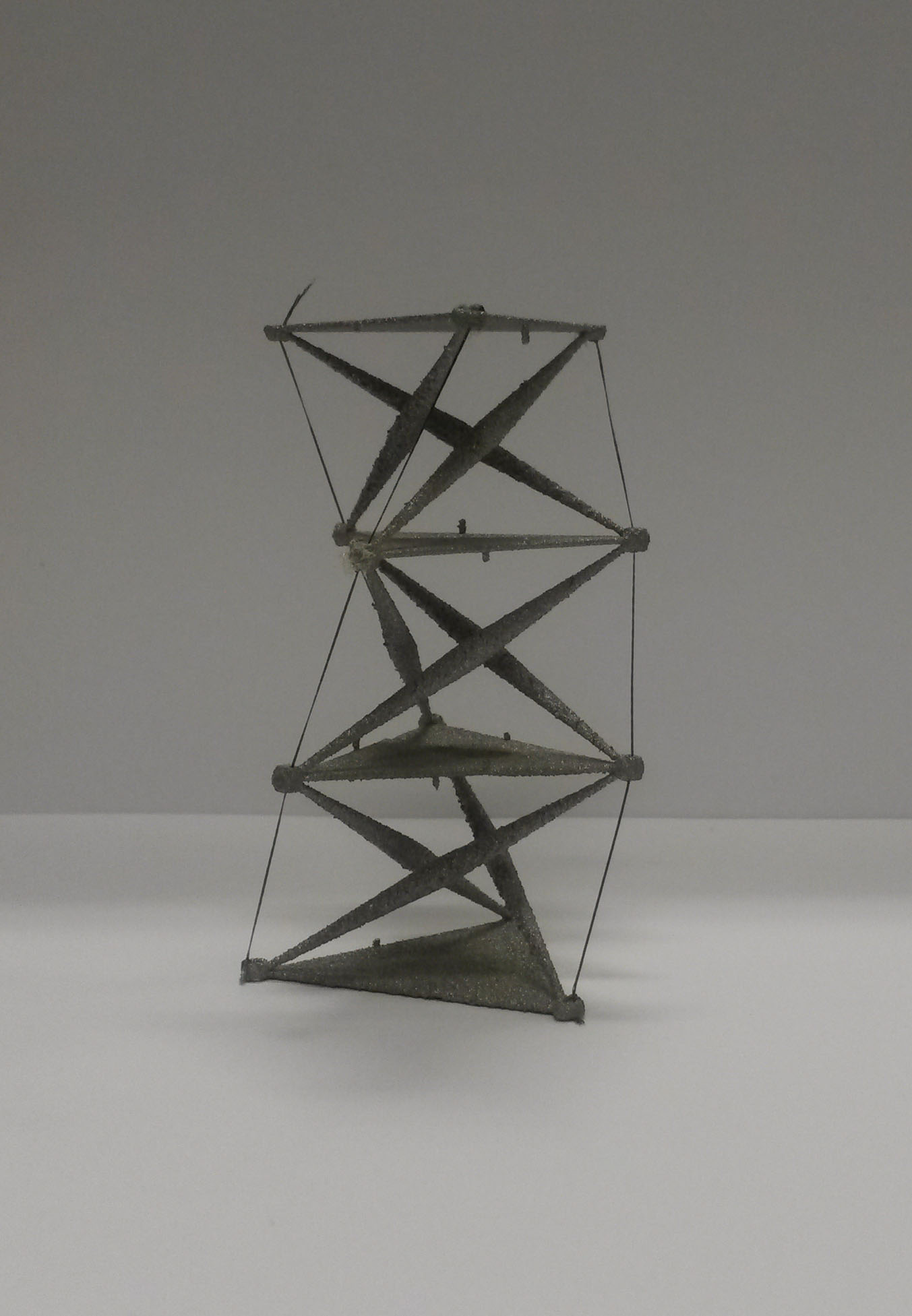}\hfil
\includegraphics[width=0.39\textwidth]{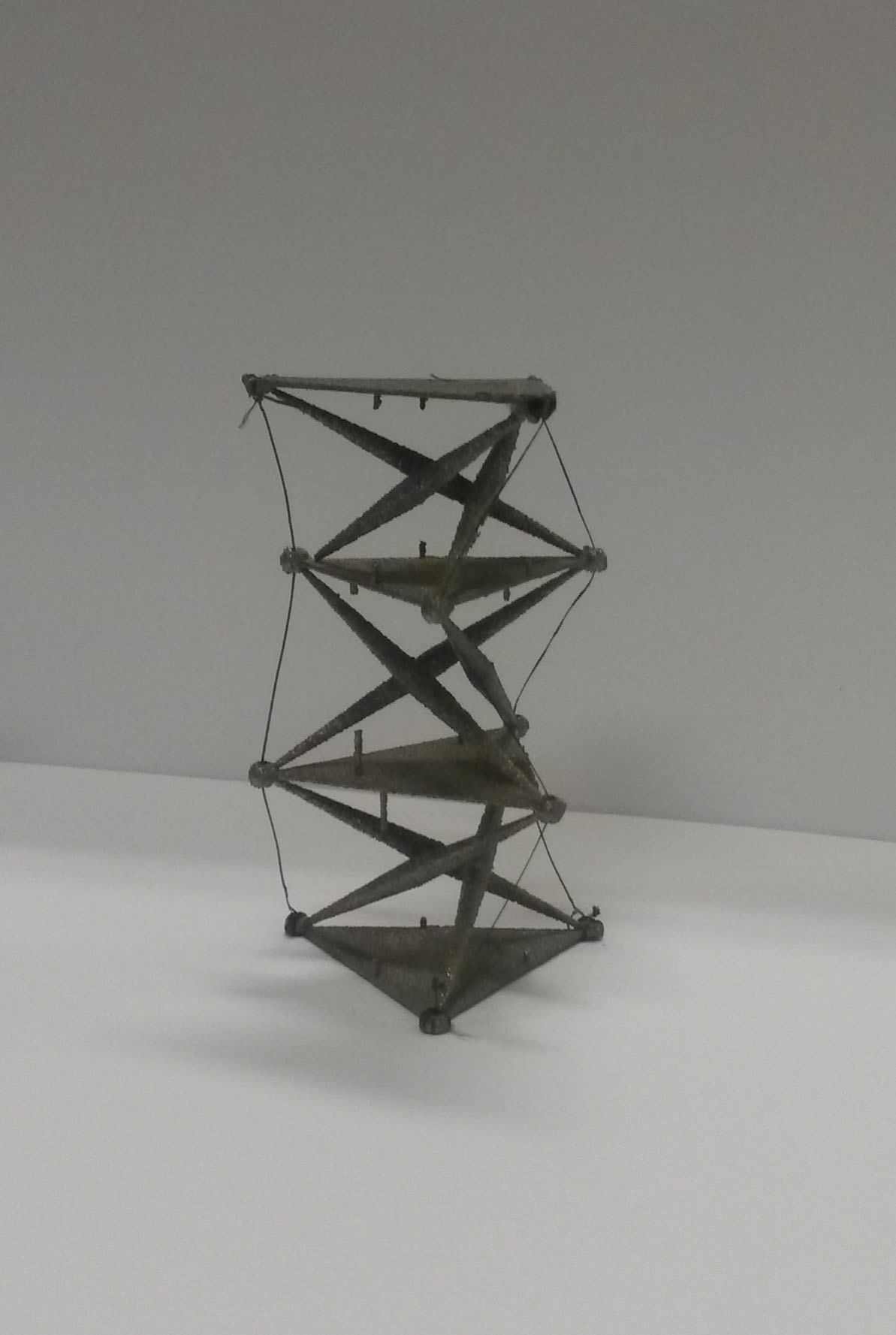}
\caption{3LLL (left) and 3LRL (right) columns obtained by `sewing' the EBM-printed models with Spectra strings.
}
\label{Sewed_columns}
\end{center}
\end{figure}

\section{Quasi-static compression tests} \label{comptest}

We investigated on the experimental response of the structures illustrated in the previous section under quasi-static compression loading, on employing a strain rate of $1 \cdot {10}^{− 3}$ $s^{−1}$. The testing apparatus consisted of a Zwick/Roell Z050 testing machine equipped with 20 kN load cell  (Fig.\ref{Experimental_setup}).
{\crev{Two steel plates were inserted between the terminal bases of the structures to be tested and the testing machine plates  (cf. Fig.\ref{Experimental_setup}).}}
To minimize the frictional effects between the terminal bases of the prisms, the steel plates and the testing machine plates, we carefully lubricated all such surfaces, before initiating the compression tests. We lubricated the joints of the tested structures as well, in order to reduce the friction between the strings and the rough internal surface of the joints.

\begin{figure}
\centering
\includegraphics[width=.50\textwidth]{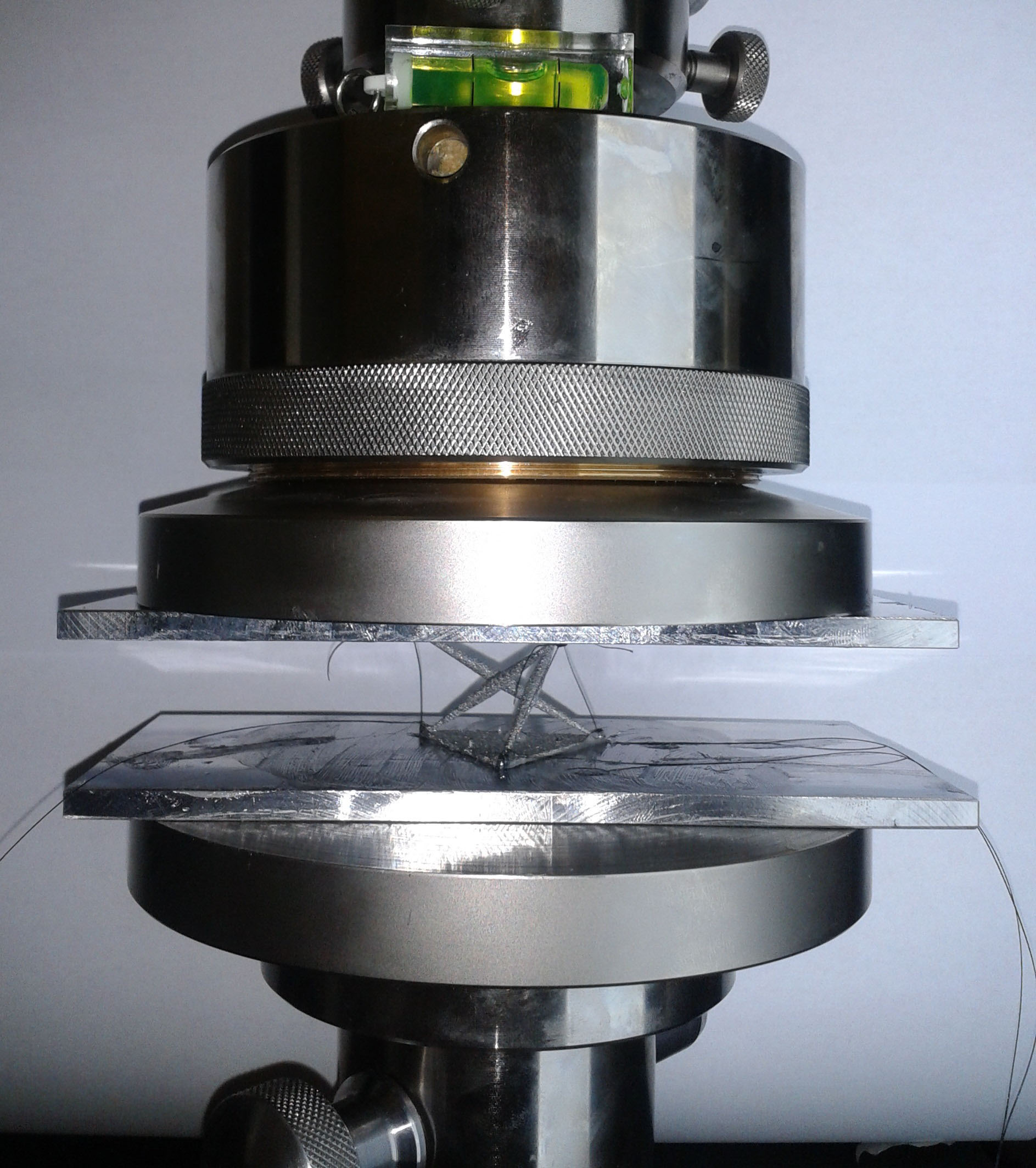}\hfil
\includegraphics[width=.447\textwidth]{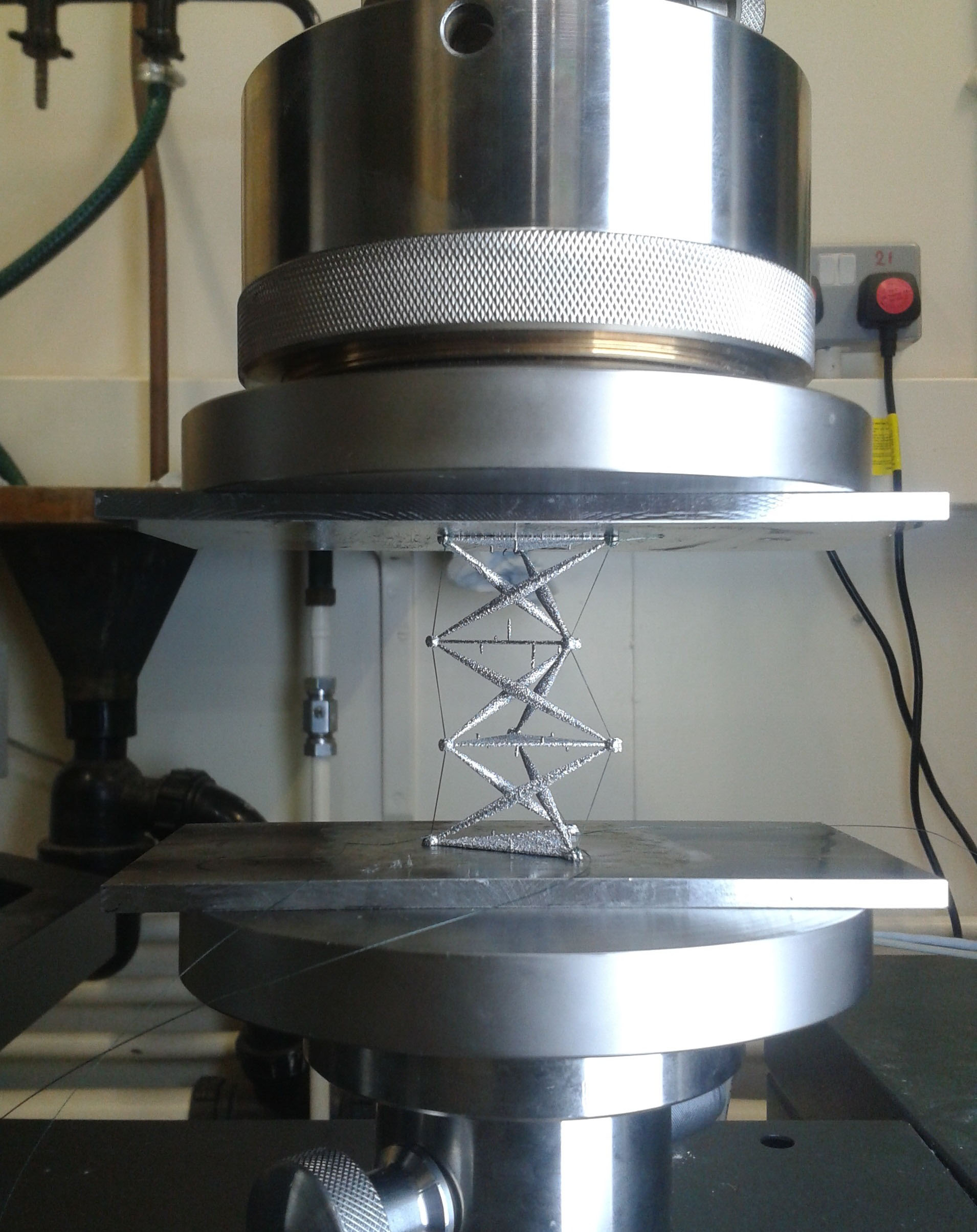}
\caption{{\cred{Experimental set-up for the compression testing of post-tensioned structures: L-prism (left), and 3LRL-column (right)}}}
\label{Experimental_setup}
\end{figure}

Fig. \ref{test_LLLprism} shows the results of compression tests on different L-prisms and R-prisms, which differ from each other for the value of the applied prestress.
Here and in the following examples, the cross-string prestrain $p$ \citep{JMPS14} was 
estimated by comparing the lengths of the cross-strings in the unstressed 
and prestressed configurations (cf. Fig. \ref{test_LLLprism}).

The axial force $F$ vs. axial-displacement $\delta$ plots in Fig. \ref{test_LLLprism} highlight a marked stiffening behavior of the single-prism structures (tangent stiffness increasing with $\delta$), up to the specimen failure.
Such an elastic stiffening (or hardening) effect is magnified by increasing the internal prestress (cf. Fig. \ref{test_LLLprism}),
in agreement with the experimental results presented in \cite{prot}, and the theoretical results given in \cite{JMPS14}, for tensegrity prisms endowed with rigid bases.
The specimen collapse was due to the detachment of one or more bars from the joints. It is worth mentioning that we did not observe plastic yielding of the strings or buckling of the bars up to the specimen failure. 

\begin{figure}[hbt] \begin{center}
\includegraphics[width=12cm]{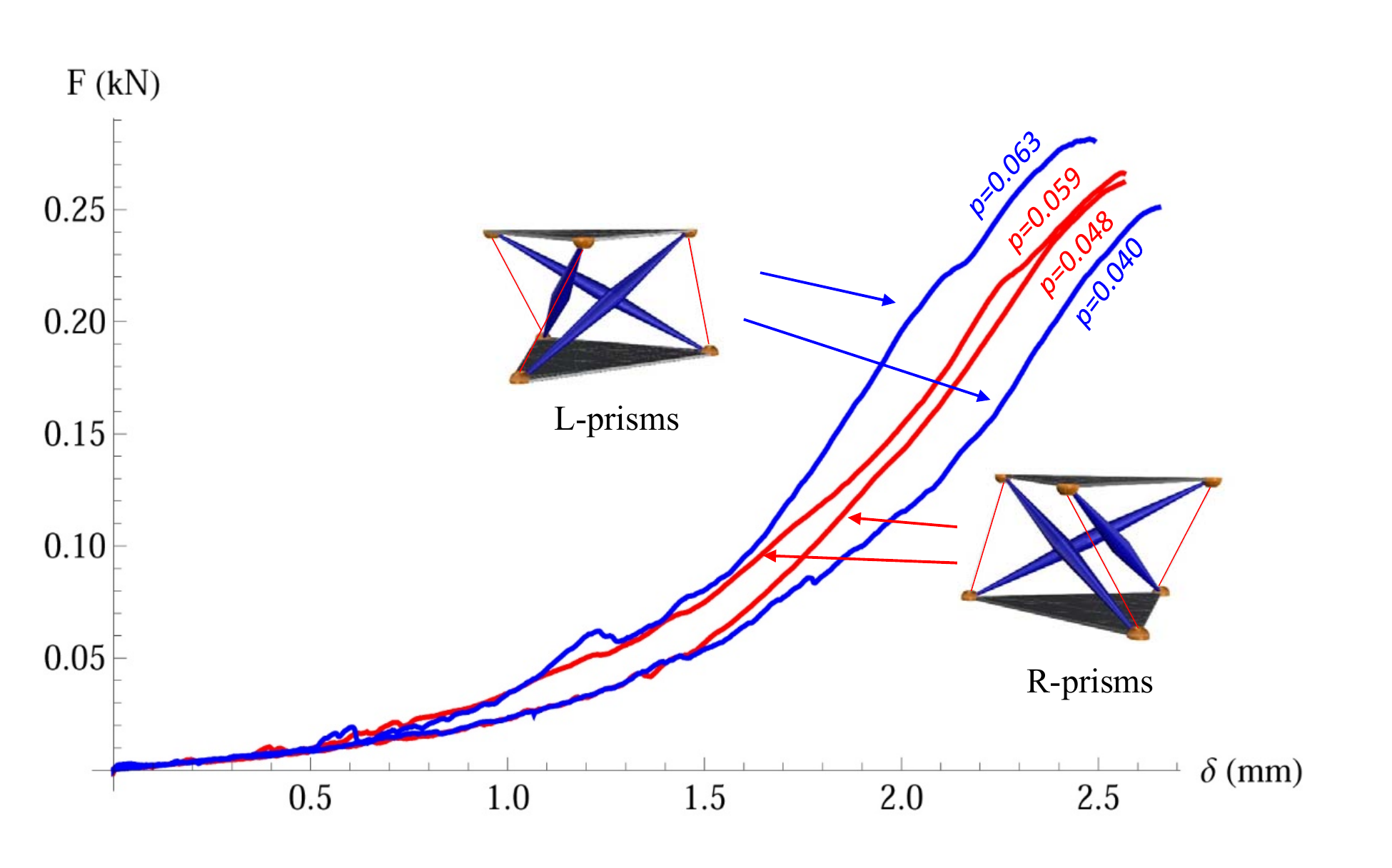}
\caption{Axial force $F$ vs. axial-displacement $\delta$ curves of single-prism structures under axial compression loads.}
\label{test_LLLprism}
\end{center}
\end{figure}

The force F vs. displacement $\delta$ curves of 3LLL and 3LRL specimens are shown in Fig. \ref{3L3LR_column}.
We observe that the initial branches of the $F-\delta$ curves exhibited by such specimens are practically coincident, for $\delta\leq 2.50$ mm.
As $\delta$ gets larger than such a threshold, we instead observe that $F - \delta$ curves of the 3LLL specimens get visibly stiffer than those exhibited by the 3LRL specimens. 
We also observe that the $F - \delta$ curves of the 3LLL specimens exhibit marked oscillations in the large displacement regime. This is explained by damage phenomena due to the marked rubbing of the cross-strings against the walls of the joint holes in such a regime.
The collapse of the 3LLL specimens was due to the detachment of the bars of the top-most prism from the upper joints, while that of the 3LRL specimens was due to the simultaneous detachment of the bars of the central prisms from upper and lower joints.
It is worth remarking that the rupture of the 3LRL specimens occurred in correspondence with smaller axial displacements $\delta$, as compared to that exhibited by the
3LLL specimens (Fig. \ref{3L3LR_column}).

The anticipated failure of 3LRL specimens can be explained by the opposite twisting of the terminal prisms, which produces marked stress concentration at the interfaces between such units and the central prism.
On the other hand, in the 3LLL specimens, all the prisms twist counter-clockwise, which produces reduced stress concentration at the interfaces between the different units, as compared to the 3LRL specimens, and allows the structure to sustain large axial displacements before failure.

Overall, we may conclude that the response of the 3LRL specimens is affected by opposite-twisting effects of the terminal units, especially under large or moderately large axial displacements, while that of the 3LLL specimens replicates the response of a single prism over a larger window of axial displacements (compare Figs. \ref{test_LLLprism} and \ref{3L3LR_column}).
In both cases, we observe a geometrically nonlinear response of the structure (more pronounced in the case of the 3LRL specimens).

\begin{figure}[hbt] \begin{center}
\includegraphics[width=14cm]{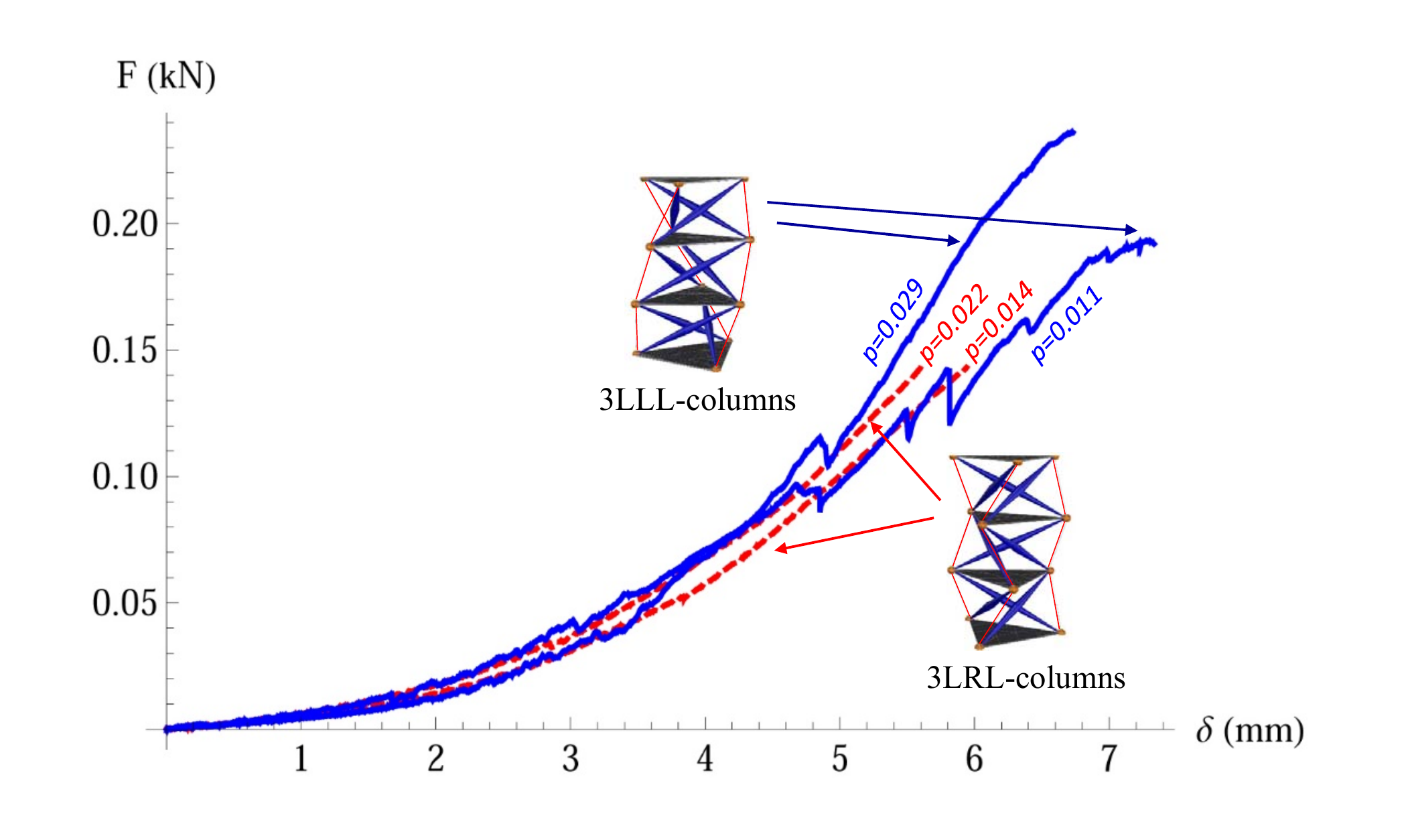}
\caption{Experimental results of the compression tests on 3LLL- and 3LRL-columns.}
\label{3L3LR_column}
\end{center}
\end{figure}

\medskip

\section{Concluding remarks} \label{conclusion}

We have investigated on the fabrication of physical models of tensegrity structures, by combining the additive manufacturing of metallic models, via EBM, with the manual insertion of tensioned Spectra strings in 3D printed structures. 
Such a `post-tensioning' approach has been applied to the construction of columns composed of different numbers of tensegrity prisms, which may feature either equal or opposite orientation \citep{Skelton2010c}. In all the examined structures, the building blocks consist of tensegrity prisms endowed with rigid bases.
Compression tests have been performed on single prisms and columns composed of three prisms, with the aim of characterizing the nature of the mechanical response of such structures under large or moderately large axial displacements.
We have observed a markedly nonlinear response of the examined structures, and elastic hardening effects under large axial displacements.
Such results  confirm the outcomes of previous theoretical and experimental studies available in the literature on the mechanical response of tensegrity prisms and columns endowed with rigid bases \citep{Oppenheim:2000, FSD12, prot, JMPS14, FCA14}.

Elastically hardening mechanical metamaterials support compressive solitary waves and the unusual reflection of waves on material interfaces \citep{Nest01,FSD12, FCA14}. Solitary wave dynamics has been proven to be useful for the construction of a variety of novel acoustic devices. These include: acoustic band gap materials; shock protector devices; acoustic lenses; and energy trapping containers, to name some examples (refer, e.g., to \cite{Theocharis2013} and references therein).
The present study represents a first step towards the additive manufacturing and testing of nonlinear tensegrity metatamerials to be employed for real-life engineering devices.
We address the formulation and implementation of innovative, multimaterial deposition techniques that are able to apply internal prestress in AM-fabricated tensegrity structures to future work.
We plan to fabricate macro- and small-scale models of such structures using materials with different coefficients of thermal expansion for struts and cables.  
In addition, we plan to manufacture micro-scale tensegrity models through projection micro-stereolitography \citep{Zheng12}, employing swelling materials for the tensile members \citep{Lee:2012}. Once dried, the tensile members will contract, creating internal prestress.
Engineering applications of the such metamaterials will deal with novel, tunable focus acoustic lenses, and innovative devices for monitoring structural health and damage detection in materials and structures \citep{FSD12, Don14,Ni11}. 

\medskip

\section*{Acknowledgements}
Support for this work was received from the
Italian Ministry of Foreign Affairs, Grant No. 00173/2014, Italy--USA  Scientific and Technological Cooperation 2014--2015
(`\textit{Lavoro realizzato con il contributo del Ministero degli Affari Esteri, Direzione Generale per la Promozione del Sistema Paese}').

\medskip

\section*{References} \label{references}

\bibliographystyle{apalike}

\end{document}